\documentclass[runningheads]{llncs}

\usepackage{array, comment, tabularx}
\usepackage{cite}
\usepackage{enumitem}
\usepackage{graphicx}
\usepackage{verbatim} 
\usepackage{url}
\usepackage{multirow}
\usepackage{booktabs,float}

\newcolumntype{Y}{>{\centering\arraybackslash}X}
\newcolumntype{y}{>{\arraybackslash}X}

\begin{document}

\title{Beyond PS-LTE: Security Model Design Framework for PPDR Operational Environment}

\author{Daegeon Kim \and Do Hyung Gu \and Huy Kang Kim}
\authorrunning{D. Kim et al.}

\institute{Graduate School of Cybersecurity, Korea University, Republic of Korea\\
\email{\{dgkim0803, iamkdh, cenda\}@korea.ac.kr}}
\maketitle              

\begin{abstract}

National disasters can threaten national security and require several organizations to integrate the functionalities to correspond to the event. Many countries are constructing a nationwide mobile communication network infrastructure to share information and promptly communicate with corresponding organizations. Public Safety Long-Term Evolution (PS-LTE) is a communication mechanism adopted in many countries to achieve such a purpose. 
Organizations can increase the efficiency of public protection and disaster relief (PPDR) operations by securely connecting the services run on their legacy networks to the PS-LTE infrastructure. This environment allows the organizations to continue facilitating the information and system functionalities provided by the legacy network. The vulnerabilities in the environment, which differ from commercial LTE, need to be resolved to connect the network securely. 
In this study, we propose a security model design framework to derive the system architecture and the security requirements targeting the restricted environment applied by certain technologies for a particular purpose. After analyzing the PPDR operation environment's characteristics under the PS-LTE infrastructure, we applied the framework to derive the security model for organizations using PPDR services operated in their legacy networks through this infrastructure. 
Although the proposed security model design framework is applied to the specific circumstance in this research, it can be generally adopted for the application environment.

\end{abstract}


\section{Introduction}
\label{sec:introduction}
Advances in mobile communication technologies enrich civilians' lives and enhance the operational efficiency for public protection and disaster relief (PPDR). The technologies connect response agents from fields and command posts to increase their field functionality. Besides, the agencies can enhance their mutual co-operation by exchanging information over a shared communication channel. 

Many countries have constructed mobile communication network infrastructures for PPDR. However, cutting-edge mobile communication technologies have rarely been adopted as a communication mechanism. Instead, some functionalities required for PPDR are added to the technologies whose reliability and safety are guaranteed through experience. 

LTE is one of the most widely used mobile communication technology and is used by 262 operators in 120 countries, the number of which continues to increase\cite{GSA2019}. The earliest standard related to LTE\footnote{3GPP, TS 36.101 (ver.1.0.0): "Evolved Universal Terrestrial Radio Access (E-UTRA); User Equipment (UE) radio transmission and reception"} was published in 2007, and the first commercial network was launched in 2010~\cite{5481625}. 

In 2013, 3GPP defined Public Safety LTE (PS-LTE) by adding the system features to LTE, namely, proximity-based service (ProSe) and group communication system enabler (GCSE). The standards of these features were included in 3GPP release 12 in 2015. One year later, in 3GPP release 13, additional PPDR functions, including mission-critical push-to-talk (MCPTT) and isolated E-UTRAN operation for public safety (IOPS), were applied.

\begin{figure}
    \centering
    \includegraphics[width=\textwidth]{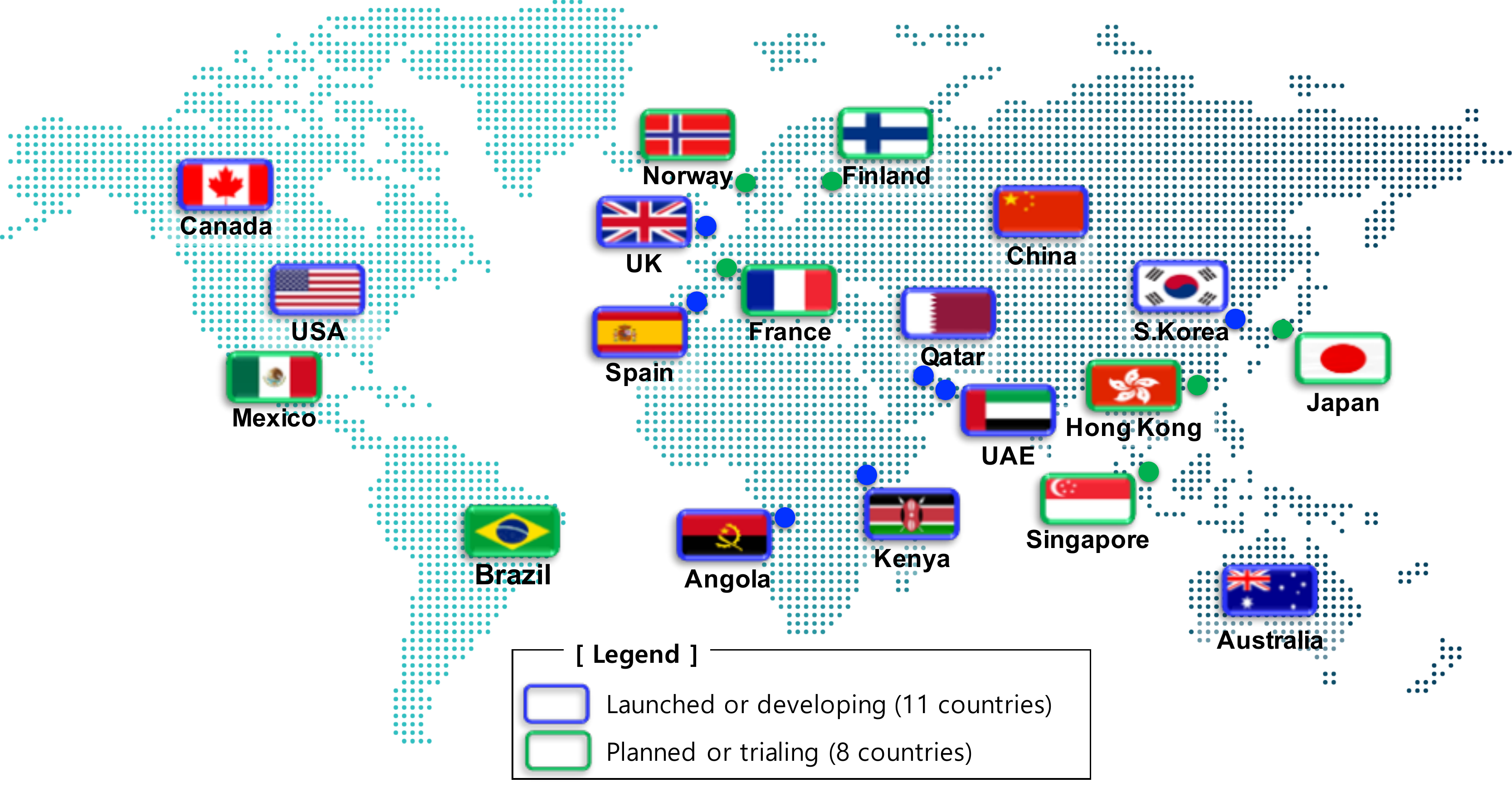}
    \caption{Global use of PS-LTE}
    \label{fig:LTE_Trends}
\end{figure}

PS-LTE networks are widely spreading as a communication mechanism for PPDR in many countries. Figure~\ref{fig:LTE_Trends} shows the global use of PS-LTE as summarized in \cite{GSA2018}\footnote{The status of the Republic of Korea updated from \textit{planned} to \textit{developing} compared to the original source.}.

Organizations using the PS-LTE infrastructure need to provide the services run on their legacy networks on the infrastructure to increase the effectiveness of a PPDR operation. The services allow the organizations to continue facilitating the information and the system functionalities provided by the legacy network. 

A security threat analysis for the operational environment is necessarily preceded when designing a secure system linking architecture between the PS-LTE infrastructure and the legacy networks. The protection mechanisms of most of the analyzed threats should be reflected in the system architecture design. However, the security requirements of the system components against other threats also need to be subtly tightened. 

We extensively consider the operational environment and the overall system elements and provide the technical guidelines practically applicable to organizations using the PS-LTE infrastructure.

\section{Contribution}
The main contributions of this paper can be summarized as follows: First, a security model design framework is used to construct an environment adapting to certain technologies for a specific purpose. In addition, we designed the system architecture and security requirements for the user organization in PPDR operational environment on PS-LTE infrastructure.

In the security model design framework section, we propose the general framework used to design the security model by analyzing the security threats, building the system architecture, and deriving the security requirements for the construction of an environment in which certain technologies are applied for a particular purpose. 

In the next section, we describe the application results of the proposed framework used by PPDR organizations under the PS-LTE infrastructure. First, we analyzed the characteristics of the operational environment from the aspect of user organization of PS-LTE infrastructure. This analysis will help the user organization raise situational awareness of their systems, applying PS-LTE infrastructure.

We constructed a test-bed where the baseline security level is applied, as described later, and conducted empirical studies of the security threats toward the test-bed environment. Finally, we designed a secure system architecture to connect a legacy network to a PS-LTE infrastructure while protecting the identified security threats. To fill in the security gaps that the architecture cannot cover and tighten the security level, we provide the security requirements of the system elements that must be satisfied. 

We want to mention that our empirical studies on the test-bed constructed using the hardware used in the PS-LTE environment in the field may not be fully generalized depending on the vendors and the system versions, the operational environment, and so on. However, the security model design framework and the application example can provide insight on how to enhance the security of PS-LTE based PPDR operational environment.

The following section provides the preliminary technical background required to understand this study. 

\section{Technical Background}
\label{sec:backgrounds}

\subsection{Basic Structure of LTE}

\begin{figure}[hbt!]
    \centering
    \includegraphics[width=0.8\textwidth]{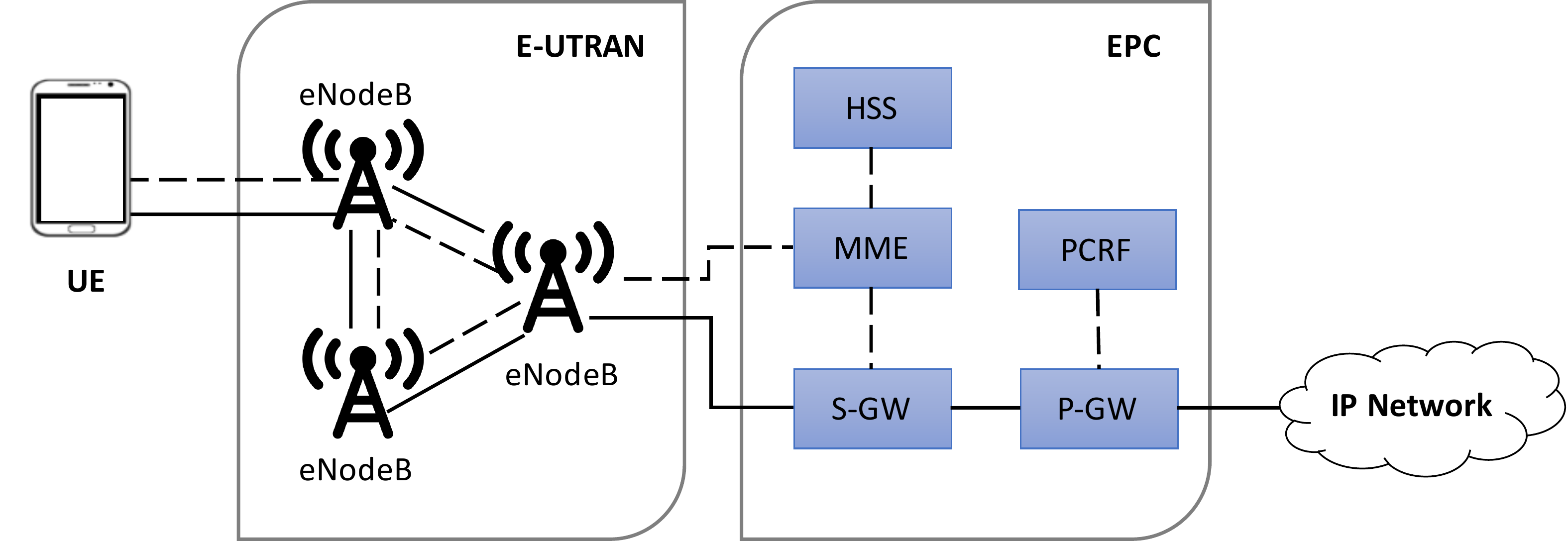}
    \caption{Basic LTE system architecture}
    \label{fig:LTE_Architecture}
\end{figure}

The basic concept of the LTE system structure is described in \cite{Cichonski2018NISTSecurity} and schematized in Figure~\ref{fig:LTE_Architecture}. In the figure, the dotted and solid lines represent two logical planes separated in an LTE network depending on the functionalities of the data; the user and the control plane, respectively. The LTE network transfers the data necessary to operate and maintain the network properly through the control plane. The user plane is responsible for carrying data that users intentionally generate (e.g., voice communication, SMS, application traffic) and send over the network. 

The following subsections describe the key points of the components making up the LTE structure. 

\subsubsection{User Equipment (UE)}
An UE is a cellular device, such as a cell phone and a tablet, interacting with base stations through radio signals. Several components make up an UE and indicate its identity as follows. 

A Mobile Equipment (ME) is a mobile terminal made by cell phone manufacturers. An Universal Integrated Circuit Card (UICC), also known as an Universal Subscriber Identity Module (USIM), is a smart card inserted into an ME. An UICC stores cryptographic keys used for the authentication and key agreement (AKA) with the LTE network. An UICC also contains a network ID, called a Public Land Mobile Network (PLMN), from which the Mobile Network Operator (MNO) of the UE obtains LTE service. 

An UE has two permanent IDs assigned by the manufacturer and the MNO, an International Mobile Equipment Identifier (IMEI), and an International Mobile Subscriber Identity (IMSI).

\subsubsection{Evolved Universal Terrestrial Radio Access Network (E-UTRAN)}
An UE transmits and receives radio signals to/from the base station, which are called Evolved Node B (eNodeB or eNB), to communicate with the core network. An E-UTRAN is a mesh network composed of eNodeBs that modulate and demodulate radio signals. 

\subsubsection{Evolved Packet Core (EPC)}
An EPC is the brain of an LTE network. It authorizes UEs to connect to the network and manages their connections. The following are the key components included within the EPC. 

\begin{itemize}[label=$\bullet$]
    \item \textbf{Mobility Management Entity (MME)}: The primary entity treats signal plane data, including the functionalities that control AKA procedures between UEs and the Home Subscriber Server (HSS). It selects gateways that route user plane data sent to/from the UEs. 
    \item \textbf{Home Subscriber Server (HSS)}: This stores the IDs and cryptographic keys matched to the UE and generates an authentication vector for the AKA.
    \item \textbf{Serving Gateway (S-GW)}: This carries user plane data and routes data between the P-GW and the E-UTRAN. 
    \item \textbf{Packet Data Network Gateway (P-GW)}: This allocates IP addresses to the UEs, routes packets, and interconnects with the PDNs. 
    \item \textbf{IP Multimedia Subsystem (IMS)}: This system provides multimedia services such as VoLTE, SMS, and MMS. 
    \item \textbf{Policy and Charging Rules Function (PCRF)}: This stores rules and policies related to the quality of service (QoS), charging, and allows network resources.
    \item \textbf{Backhaul}: This links the E-UTRAN (eNodeBs) and the EPC.
\end{itemize}

\subsubsection{IP Network}
Any external IP network connected to an LTE network is called a Packet Data Network (PDN). The P-GW routes the data from a PDN. The gateway between the EPC and PDN is called an Access Point Name (APN), which also serves as the identifier of the PDN. An UE must be assigned APNs to connect to the PDNs.

\subsection{PS-LTE Characteristics}
PS-LTE provides additional features to support an effective PPDR operation compared to a conventional LTE system. Group communication system enabler (GCSE) is a fast and efficient mechanism to distribute various media content to multiple users in a controlled manner. Proximity-based Services (ProSe) enable direct and relayed communications among neighboring UEs without passing the core network. IOPS provides the ability to maintain communications following the loss of a backhaul connection. In addition, MCPTT supports enhanced PTT services suitable for mission-critical scenarios. 

Table~\ref{tab:PS-LTE_Docs} lists above features and related 3GPP documents. The specifications of the features can be found in the documents. 

\begin{table}[H]
\caption{PS-LTE specific functionalities and corresponding 3GPP documents (TS, technical specification; TR, technical report)}
\label{tab:PS-LTE_Docs}
\centering
\begin{tabular}{ccccc}
\toprule
\textbf{Functionality}	& \textbf{Abbreviation}	& \textbf{\begin{tabular}[c]{@{}c@{}}Requirements\\ (function/security)\end{tabular}} & \textbf{Architecture} & \textbf{\begin{tabular}[c]{@{}c@{}}Security\\ issues\end{tabular}}\\
\midrule
\begin{tabular}[c]{@{}c@{}}Group Communication\\ System Enablers\end{tabular} & GCSE & TS 22.468 & TR 23.768 & TS 33.888\\
Proximity-based Services & ProSe & \begin{tabular}[c]{@{}c@{}}TS 22.803\\ TS 22.278\end{tabular} & TR 23.703 & \begin{tabular}[c]{@{}c@{}}TS 33.303\\ TS 33.833\\ TS 33.843\end{tabular}\\
\begin{tabular}[c]{@{}c@{}}Isolated E-UTRAN Operation\\ for Public Security\end{tabular} & IOPS & \begin{tabular}[c]{@{}c@{}}TS 22.346\\ TS 22.897\end{tabular}              & -            & TS 33.897\\
Mission-Critical Push-To-Talk                                                            & MCPTT & \begin{tabular}[c]{@{}c@{}}TS 22.179\\ TS 22.280\end{tabular}              & TR 23.779    & \begin{tabular}[c]{@{}c@{}}TS 33.179\\ TS 33.879\\ TS 22.280\end{tabular} \\
\bottomrule
\end{tabular}
\end{table}

\section{Related Studies}
\label{sec:RelatedWork}

Researches into LTE security vulnerabilities help to understand security threats on the proposed security model design framework.
In \cite{Hussain2018LTEInspectorAS}, LTEInspector is proposed, which analyzes the LTE system by leveraging the combined power of a symbolic model checker and a protocol verifier through a model-based adversarial testing approach. For 3 critical procedures of the 4G LTE protocol (attach, paging, and detach), 10 novel and 9 known attacks were found using LTEInspector. 
 
In \cite{8835335}, the vulnerabilities of the RRC protocol are analyzed corresponding to the layer two LTE protocol. The authors assumed 2 types of attack models, passive and active. In their analysis, they found identity mapping and website fingerprinting vulnerabilities under the passive attack model. In addition, a DNS spoofing vulnerability was identified through the passive attack model. The authors also demonstrated the feasibility of all 3 attacks using realistic setups. 

In \cite{Kim2019TouchingPlane}, a semi-automated testing tool, LTEFuzz, is implemented, which is a dynamic testing tool targeting the control plane procedures of an LTE network. The authors identified 15 known and 36 new vulnerabilities among different commercial LTE networks and device vendors. They also demonstrated several attacks based on the vulnerabilities. The attacks caused a denial of service, phishing messages, and eavesdropping/manipulation of the data traffic. 

Table~\ref{tab:RelatedWork} summarizes the characteristics of the approaches in \cite{Hussain2018LTEInspectorAS, 8835335, Kim2019TouchingPlane}.

\begin{table}[t]
\caption{Studies on LTE protocol security}
\label{tab:RelatedWork}
\centering
\begin{tabular}[t]{m{2cm}*{3}{>{\raggedright\arraybackslash}p{0.24\linewidth}}}
\toprule
&      \textbf{LTEInspector}\cite{Hussain2018LTEInspectorAS} & \textbf{LTE Layer 2 Vul.}\cite{8835335} & \textbf{LTEFuzz}\cite{Kim2019TouchingPlane}\\ \midrule
\begin{tabular}[c]{@{}l@{}} analysis target \\ protocol \end{tabular} & NAS (attach, paging, detach) & MAC and PDCP & RRC and NAS \\
\midrule
approach & \hangindent=1.5mm- model based adversarial approach & \hangindent=1.5mm- manual analysis & \hangindent=1.5mm- semi-automatic fuzzing \\
\midrule
\multirow{2}{*}{pre-requisite} & \hangindent=1.5mm- adversarial model construction & {[}fingerprinting attack{]} websites metadata acquisition & \hangindent=1.5mm- building pack DB for fuzzing \\ 
& security property definition  & {[}aLTEr attack{]} deleting victim's DNS and http cache & \hangindent=1.5mm- targeting TMSI acquisition   \\
\midrule
\multirow{3}{*}{cons} & \multirow{3}{\linewidth}{manually prepare the prerequisites} & \hangindent=1.5mm- proposing the results in experimental environment & \hangindent=1.5mm- no protocol interaction for fuzzing                                        \\
                      &                                                     & \hangindent=1.5mm- {[}fingerprinting attack{]} low website coverage  & \multirow{2}{\linewidth}{\hangindent=1.5mm- generated fuzzing message simply by packet field mutation} \\
                      &                                                     & \hangindent=1.5mm- {[}aLTEr attack{]} unrealistic pre-requisite      & \\
\bottomrule
\end{tabular}
\end{table}

In \cite{4437813}, the researchers survey a number of new security threats to cause unexpected service interruption and disclosure of information in 4G. They also found there still remain several open issues although many are working on fixing and/or designing new security architectures for 4G. This helps us to build security model design.

There are security threats on the LTE system in \cite{7547270, securityfor4g5g, 7452266}. These works are relevant to our security threat analysis on \ref{sec:ThreatAnalysis}. We analyze the security threats on PPDR operational environment and build system architecture preventing security threats.

In \cite{securityfor4g5g}, the researchers survey existing authentication and privacy-preserving schemes. They present four threat models classified into privacy, integrity, availability, and authentication, three countermeasures classified into cryptography methods, humans factors, and intrusion detection methods. They provide a taxonomy and comparison of authentication and privacy-preserving schemes for 4G and 5G cellular networks in the form of tables.

The LTE security threats against jamming, spoofing, and sniffing at physical channels are researched in \cite{7452266}. The researchers measured each LTE jamming attack's complexity and efficiency and identified which channel/signal is the weakest. Due to LTE is not designed to become a mission-critical communication technology, it is highly vulnerable to jamming attacks.

In \cite{7547270}, attacks toward the LTE system are classified into four groups: 1) attacks against security and confidentiality such as Evolved Packet System Authentication and Key Agreement (EPS-AKA) security issues or a management handover key failure, 2) IP-based attacks against a backhaul, GPRS Tunneling Protocol (GTP), voice over LTE (VoLTE) Session Initiation Protocol (SIP), and diameter, 3) attacks on the signal plane and 4) jamming attacks on the physical layer. 

The IMSI-catchers, also known as cell-site simulators or stingrays, are threats to LTE system subscribers. They act as rogue base stations that can track cellphone locations and often eavesdrop on cellular communications. The works for catching IMSI-catchers help us to analyze threats against rogue base stations.

In \cite{imsicatcher}, there are two implementations of the IMSI Catcher Catcher (ICC). IMSI Catchers identify and eavesdrop on phones in mobile networks, and ICC detects this threat. They implemented the ICC with stationary measurement units and app for standard consumer grade mobile phones.

In \cite{seaglass}, SeaGlass is a city-wide cell-site simulator detector. SeaGlass is capable of detecting anomalies across a wide variety of signature classes, potentially caused by actual cell-site simulators. This may be needed to PPDR operational environment preventing from city-wide tracking and eavesdropping on
cell phones.

The other proposal for enhancing subscribers' security is using multiple IMSIs for a mobile telephony subscriber. The proposed schemes in \cite{improveprivacy} provide a form of pseudonymity on the air interface, even when it is necessary to send the IMSI in cleartext. The schemes reduce the impact of user privacy threats arising from IMSI capture.

Although the LTE system is designed to be secure, threats maybe still existing. The existing errors in implementation or configuration generate threats to the system. The LTE specification must be implemented accurately in PPDR operational environment.

In \cite{10.1145/3317549.3324927}, 4 misconfigured commercial networks and multiple cases of implementation issues are reported. The researchers analyze the security configuration and test the security algorithm selection in a total of 12 LTE networks in 5 European countries. 

According to \cite{Rupprecht2016PuttingLS}, several modern smartphones are not implemented with the LTE specification. They do not inform the user that even the user data is sent unencrypted. The researchers present Man-in-the-Middle (MitM) attack against an LTE device that does not fulfill the network authentication requirements.

The srsLTE, in \cite{srsLTE}, is an open-source platform for LTE experimentation, designed for maximum modularity and code reuse and fully compliant with LTE Release 8. It is applicable to experimental LTE test-bed platforms and testing LTE configuration or implementation. It can be used as either UE or base stations with software-defined radio (SDR) device. This implementation helps us to understanding attacks by UEs (\ref{sec:category1}) like type 2 and type3 or rogue base station (\ref{sec:category2}) in our research.

In \cite{gutirealloc}, among 28 carriers, 19 carriers have easily predictable and consistent patterns in GUTI reallocation mechanisms. Revisiting 4 carriers, they also have predictable patterns after invoking GUTI reallocation multiple times within a short time period. By using this predictability, the adversary can track subscribers' locations.

Early VoLTE implementations contain several vulnerabilities that lead to serious exploits, such as caller spoofing, over-billing, and denial-of-service attacks. VoLTE is also used in mission-critical-push-to-talk (MCPTT), which is one of functionality in PS-LTE. \cite{Kim2015BreakingAF} and \cite{10.1145/2810103.2813618} are dealing with these vulnerabilities in VoLTE.

Unlike the traditional call setup, the VoLTE call setup is controlled and performed at the Application Processor (AP), using the SIP over IP. A legitimate user who has control over the AP can potentially control and exploit the call setup process to establish a VoLTE. In \cite{Kim2015BreakingAF}, the researchers identified a number of vulnerabilities of early VoLTE and proposed immediate countermeasures that can be employed to alleviate the problems but the more comprehensive solution that eliminates the root causes may be needed.

In \cite{10.1145/2810103.2813618}, several vulnerabilities exist in both control-plane and data-plane functions that can be exploited to disrupt both data and voice in operational networks. The proof of concept attacks are validated using commodity smartphones in two Tier-I US mobile carriers. It is possible that these vulnerabilities also exist in smartphones used in PS-LTE system.

Four root causes for attacks in the current mobile network (2G, 3G, and 4G) are analyzed in \cite{srtowardsfm}: wireless channel, protocol context discrepancy, an implementation issue, and specification issue. The researchers categorize known attacks by their aim, proposed defenses, underlying cases, and root causes. This paper classifies threats into root causes compared to our paper.

The authentication and key agreement (AKA) algorithm used in LTE system has several vulnerabilities. \cite{secenhancedauth, saforLTE} are researches about identified vulnerabilities of AKA and improved authentication algorithm. These are needed for our security requirements to UE and PS-LTE infrastructure (\ref{sec:SysArchitecture}).

In \cite{secenhancedauth}, Evolved Packet System Authentication and Key Agreement (EPS AKA) procedure is used to provide mutual authentication between the user and the network in the LTE/SAE architecture has several vulnerabilities such as disclosure of user identity, MitM attack. The proposed Security Enhanced EPS AKA (SE-EPS AKA) can satisfy the security and efficiency properties in the LTE/SAE architecture.

In \cite{saforLTE}, the lack of identity protection at the first initial attaches and the lack of perfect forward secrecy for the AKA mechanism are access-level security issues that may arise at the eNodeB, UE and MME level. The proposed usage of Password-Authenticated Key Exchange by Juggling (J-PAKE) protocol instead of AKA protocol suited for use in the mobile device environment.

We analyze security threats and design the system architecture (\ref{sec:SysArchitecture}), enhancing security. The researches about security requirements and LTE security enhancement helps us to propose system security requirements in detail.

In \cite{sreforcps}, eight main Security Requirement Engineering (SRE) activities are proposed for Cyber-Physical System (CPS). The purpose of these activities is to identify security requirements in a heterogeneous CPS system. In the case study of smart car parking systems, 40 security requirements are elicited following their activities. Compared to our research, this work just focused on the efficiency of the SRE framework. The researchers identified security threats and assessed the risks of a car parking system to evaluate eight SRE frameworks.

In \cite{zigbeeiot}, the proposed Security Improvement Framework (SIF) can predict and protect various potential malicious attacks in the Zigbee network and respond accordingly through a notification to the system administrator. The designed SIF has been implemented in an office security system as a case study for real-time monitoring. The evaluation results show the capacity for detecting and protecting several potential security attacks. The researchers have categorized attacks by key requirements and network layers. There are some limitations to applying this methodology to our works because our study's target system is more complicated than the Zigbee network.

\section{Security Model Design Framework}
\label{sec:framework}
 Some systems may be used in ways that were not intended at the time of development. In addition, some technologies are adopted to implement environments that are not intended to be developed. In either case, the security model should be properly designed for modified environments. 

\begin{figure}[hbt!]
    \centering
    \includegraphics[width=\textwidth]{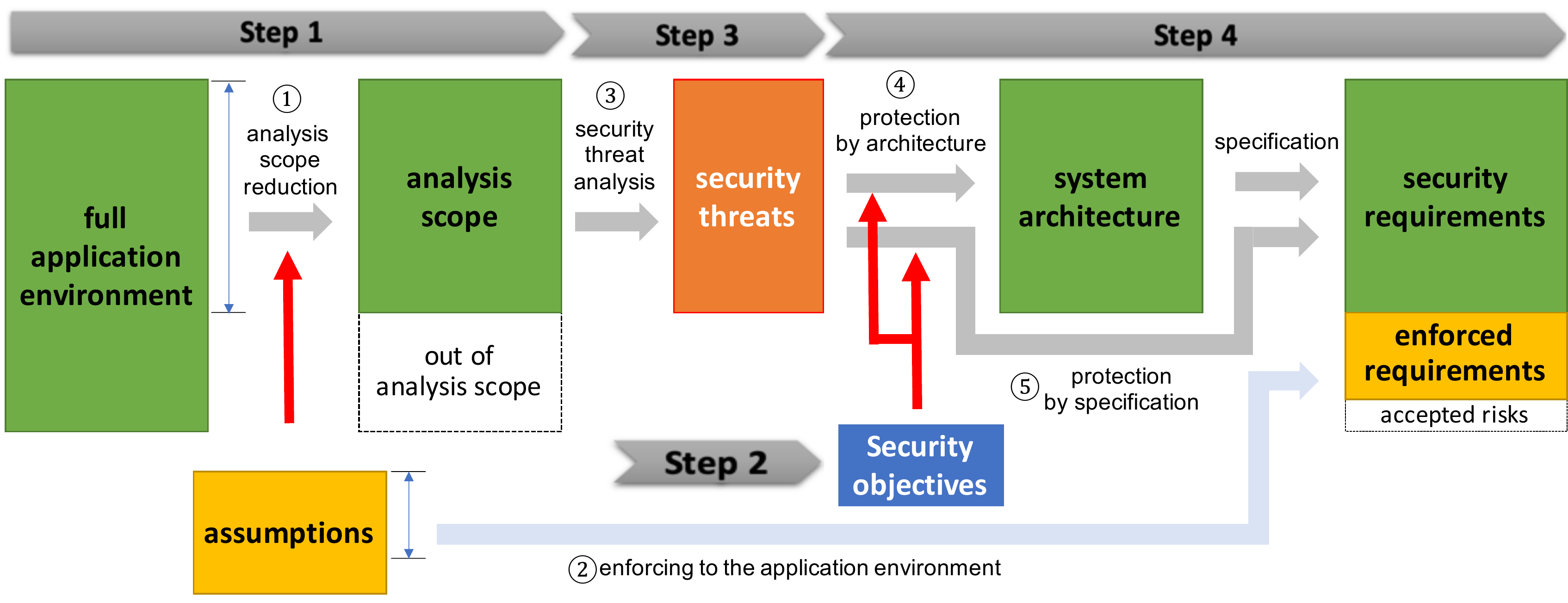}
    \caption{Security model design framework}
    \label{fig:DesignFramework}
\end{figure}

To overcome this issue, we propose a practical framework for the security model design for a particular application environment, as shown in Figure~\ref{fig:DesignFramework}. The framework enables the design of the security model for the system composed of the components developed as heterogeneous purposes.

To reduce the scope of a security threat analysis (\textcircled{\small{1}}), our framework applies certain assumptions based on the application environment and the actual restrictions. If necessary, some of the assumptions can be satisfied by enforcing them as the security requirements (\textcircled{\small{2}}). The threats neither within the analysis scope nor covered by the assumptions are accepted risks. Following the security analysis for the reduced analysis scope (\textcircled{\small{3}}), some of the threats are protected by modifying the system architecture (\textcircled{\small{4}}), and others by specifying the security requirements of the system (\textcircled{\small{5}}). 

In this framework, one can adopt any available security threat modeling methods, including those introduced in \cite{shevchenko2018threat}. If the method requires the data flow diagram (DFD) as an input, the coverage of the DFD should be restricted by applying the analysis scope.

The goal of the security model design framework is to improve the security of a system. Some of the threat modeling methods also provide a guide to discover security controls that effectively remove, counter, or mitigate all relevant vulnerabilities. For example, PASTA~\cite{ucedavelez2015risk} includes the countermeasure indication process. Since PASTA focuses on the software security aspect, the countermeasures are derived as the form of additional security functions. LINDDUN~\cite{wuyts2015linddun} deals with security problems as the privacy aspect. In the mitigation strategy elicitation step of LINDDUN method, the privacy-enhancing technologies (PETs) are provided to obtain privacy. OCTAVE~\cite{caralli2007introducing} has the step to select the protection strategy among \textit{accept}, \textit{mitigate}, and \textit{defer} as introduced in ISO 20071.
These threat modeling methods provide conceptional mitigation strategies, techniques, and functionalities. Compared with these threat model methods, our security model design framework specifies the mitigation strategy to the system architecture modification and the security requirements specification, which includes additional functionalities and software modifications. Therefore, this framework helps to understand how to reflect the mitigation strategy to the system.

The framework can be clarified through the application demonstrated in the following section. As aforementioned, the mitigated strategy toward the security threats belonging to Table~\ref{tab:AnaysisScope} against the initial system in Figure~\ref{fig:AnaysisScope} are specified to the improved system structure in Figure~\ref{fig:SystemArchitecture} and to the system requirements in Table~\ref{tab:SystemRequirements}.

\section{Application to the PPDR Operational Environment under PS-LTE Infrastructure}
\label{sec:Application}

In this section, we demonstrate the application example of the proposed framework on the PS-LTE infrastructure, which is used for the PPDR operational environment.

We found that UE included the most threats, and the threats causing high-level impacts were included in EPC. At the end of the section, the security model to mitigate the threats is provided.

Before the scope restriction step, the operational environment characteristics from the user organization aspect of the PS-LTE infrastructure need to be analyzed to support the situation awareness.

\subsection{Analysis of Operational Environment}
\label{sec:EnvAnalysis}
PS-LTE is a network infrastructure allowing the PPDR organizations to communicate and share information regarding PPDR operations. To conduct security threat analysis and derive proper security requirements, it is crucial to understand the operational environment through which several organizations share infrastructure and connect their legacy systems. 

\begin{figure}[hbt!]
    \centering
    \includegraphics[width=0.8\textwidth]{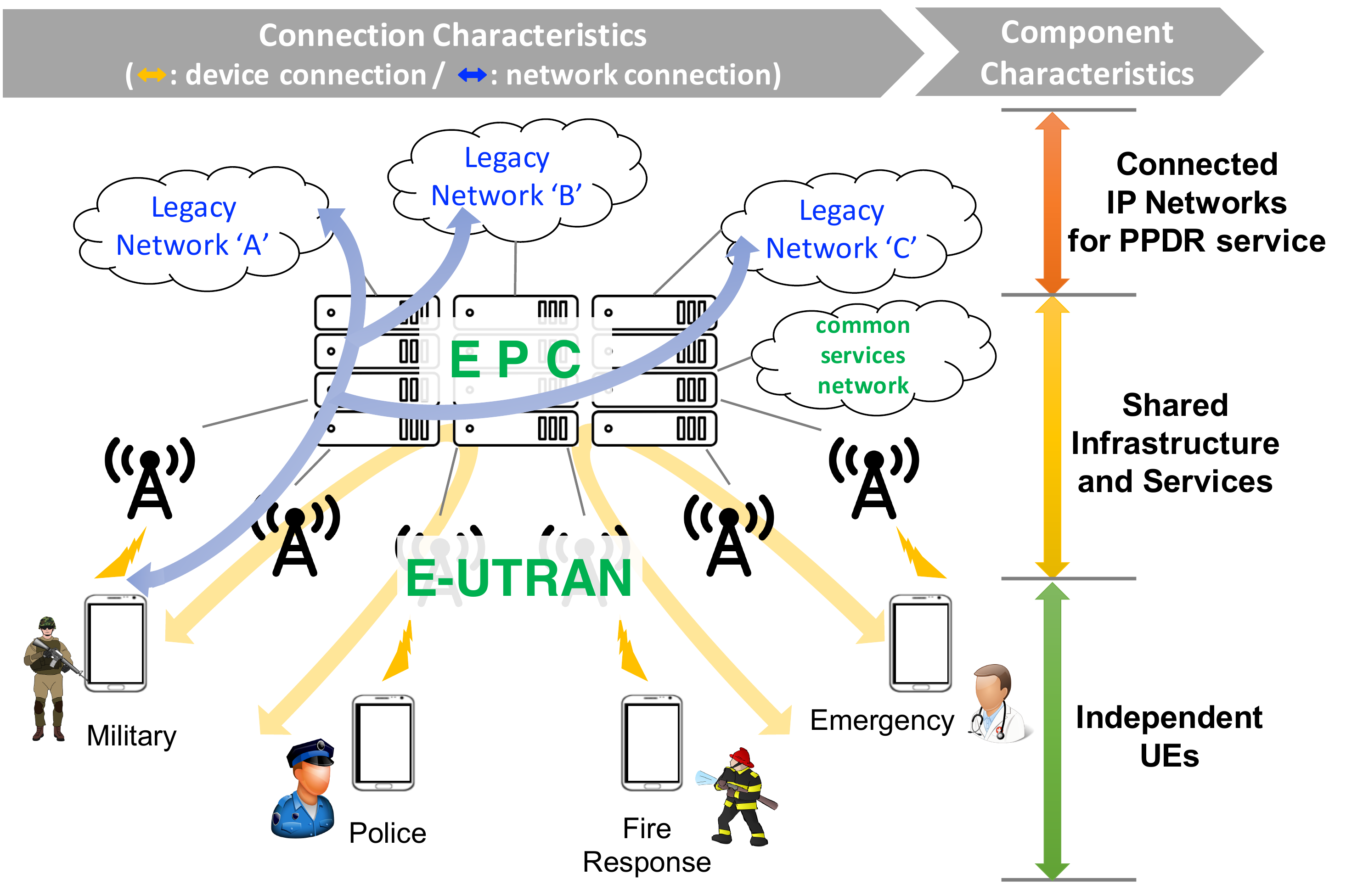}
    \caption{PPDR operational environment characteristics based on the PS-LTE}
    \label{fig:Op_Environment}
\end{figure}

Figure~\ref{fig:Op_Environment} demonstrates the characteristics of the PPDR operational environment based on the PS-LTE infrastructure. 

The components consisting of the operational environment can be categorized into UEs, the LTE infrastructure, and IP networks. Individual users utilize the UEs owned by their organizations, and the device information is registered in the LTE infrastructure, specifically in the HSS. All personnel conduct operations using their UEs under the shared LTE infrastructure (eNodeBs and EPC), and services (e.g., VoLTE, SMS, and MCPTT) are managed and controlled by another authority. Legacy IP networks can be connected to the LTE infrastructure to provide unique services required for each organization. 

In terms of the connection characteristics, UEs can communicate not only with UEs belonging to the same organization but also with those belonging to different organizations. The connected legacy IP networks are reachable from all registered UEs even when they belong to different organizations. 

These characteristics invoke environment-specific vulnerabilities, which must be prevented using features primarily provided by the LTE system.

\subsection{Assumptions}
\label{sec:Assumptions}
Two assumptions and their effects on the analysis are described below. 

\subsubsection{A1. PPDR organizations are unable to affect a PS-LTE system.}
The requirements of the PS-LTE system are defined in the standards, as summarized in Table~\ref{tab:PS-LTE_Docs}. Because PS-LTE is based on the LTE system, more standards exist to define a plain LTE system. Although several vulnerabilities caused by the standard issues have been reported \cite{Kim2019TouchingPlane}, PPDR organizations are typically not the stakeholders resolving such issues. 

By this assumption, we exclude the LTE and PS-LTE standards from our research scope.

\subsubsection{A2. Security of the shared infrastructure and services are provided by the host organization.}
The authority and responsibility to maintain and control the shared infrastructure and common services, as shown in Figure~\ref{fig:Op_Environment}, are typically established for an organization. Security requirements of the infrastructure and services should be applied and verified during the system construction. In addition, they should be monitored by the organization. The PPDR organizations need to trust the security status maintained by the host organization. 

Based on this assumption, we consider the threats to the UEs, and the IP network, and the threats from inside attackers who are authorized to connect to the EPC and rogue eNodeBs that are not linked to the infrastructure. 

\subsection{Analysis Scope Reduction}
\label{sec:AnalScope}
\begin{figure}[hbt!]
    \centering
    \includegraphics[width=0.95\textwidth]{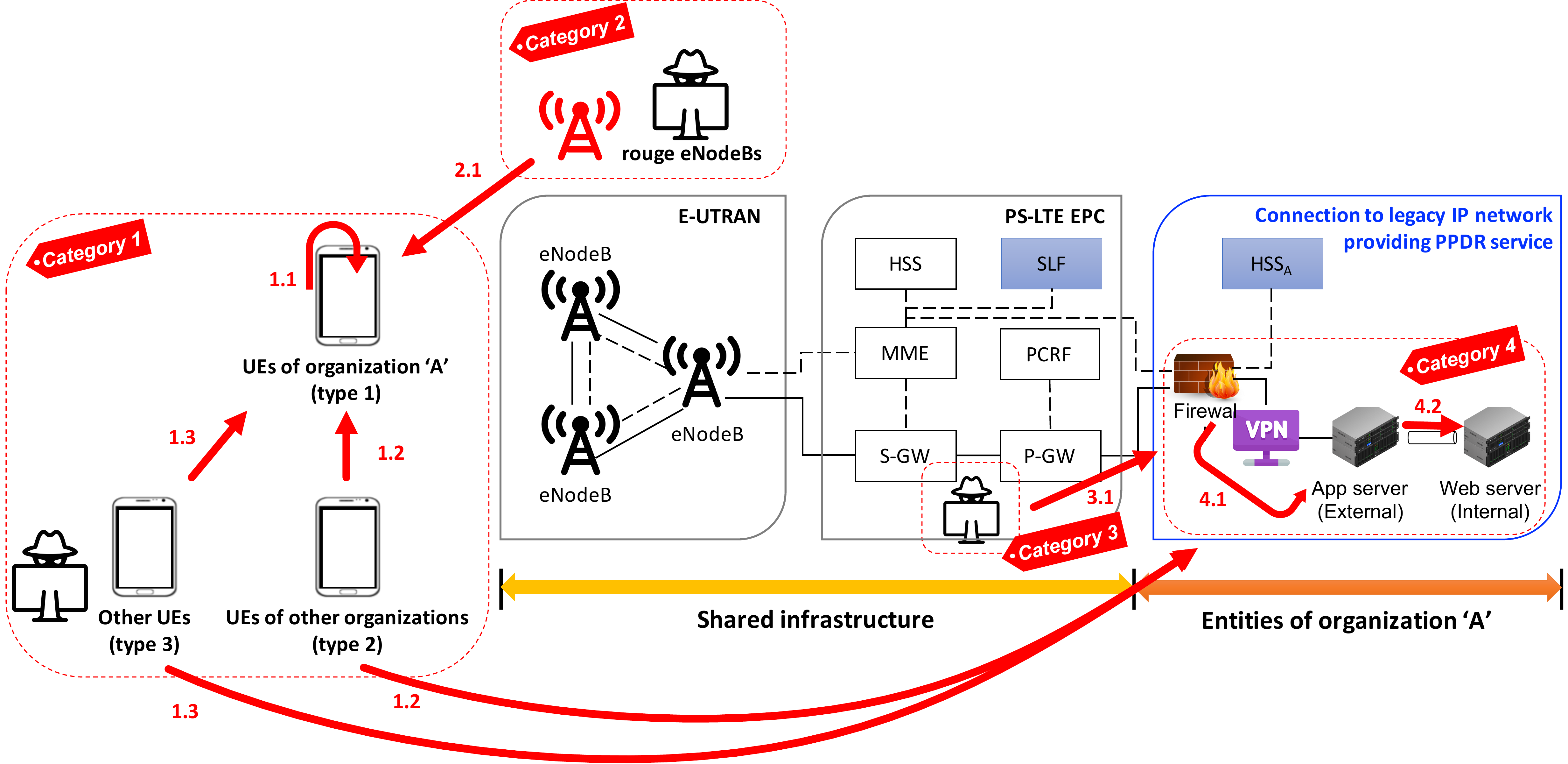}
    \caption{Categories of security threats and test-bed structure applied baseline security}
    \label{fig:AnaysisScope}
\end{figure}

We categorized the types of security threats within the research scope considered in this study. To analyze the security threats and conduct empirical studies on them, we also designed and built a test-bed. The threat categories and the test-bed structure are graphically shown in Figure~\ref{fig:AnaysisScope}.

\subsubsection{Category 1. UE}
\label{sec:category1}
UEs are sub-categorized into three types depending on their authorization to use the infrastructure and entities. A type 1 UE is owned by the same organization operating the linked entities and is able to obtain services provided by these entities. Thus, a type 1 UE is authorized to use the infrastructure and linked entities. A type 2 UE is owned by other organizations using a shared infrastructure. This type of UE can use the infrastructure but should be prohibited from using the linked entities. A type 3 UE is not even authorized to use the infrastructure regardless of having the transmission ability of the same physical radio frequency as type 1 and 2 UEs. 

Each type of UE comprises a security threat. Type 1 UEs can be misused (1.1), and type 2 and 3 UEs can threaten type 1 UEs (1.2 and 1.3). Furthermore, all types of UEs can have an adversarial effect on the connected entities (1.4).

\subsubsection{Category 2. eNodeB}
\label{sec:category2}
Assumption \textit{A1} excludes eNodeBs belonging to the shared infrastructure from this research. However, eNodeBs that are not connected to the infrastructure for an adversarial intention still threaten type 1 UEs by resulting in an unintended connection (2.1).

\subsubsection{Category 3. EPC}
Based on assumption \textit{A2}, the threats through the EPC are monitored, although it is difficult to monitor those originating from the EPC. We consider the threats by an insider attack (3.1) from the EPC side to the IP networks and the HSS\textsubscript{A}, which is operated by the user organization and set as baseline security. The reason for this is described later.

\subsubsection{Category 4. IP network}
We apply a virtual private network (VPN) gateway and an external app server in the baseline security of the IP network, the reasons for which are described later. We also derive the threats that can bypass the baseline security (4.1 and 4.2).
 
 \hfill \break
Table~\ref{tab:AnaysisScope} summarizes the categories of the security threats within this research scope.

\begin{table}[H]
\caption{Categories of security threats within the research scope}
\label{tab:AnaysisScope}
\centering
\begin{tabular}{cccl}
\toprule
\textbf{No.}	& \textbf{Category}	& \textbf{Label} & \textbf{Explanation}\\
\midrule
\multirow{3}{*}{1} & \multirow{3}{*}{UE}         & 1.1   & Threats of type 1 UEs                              \\ 
                   &                             & 1.2   & Threats from type 2 to type 1 UEs and to linked IP networks         \\  
                   &                             & 1.3   & Threats from type 3 to type 1 UEs and to linked IP networks             \\ \midrule
2                  & eNodeB                      & 2.1   & Threats from rogue eNodeBs                         \\ \midrule
3                  & EPC                         & 3.1   & Threats caused by insider attack from EPC          \\ \midrule
\multirow{2}{*}{4} & \multirow{2}{*}{IP network} & 4.1   & Threats bypassing the transfer ciphering (VPN gateway)                      \\ 
                   &                             & 4.2   & Threats from external app server   \\
\bottomrule
\end{tabular}
\end{table}

\subsection{Test-bed structure}
\label{sec:testbedStructure}

To conduct an empirical study and analyze the security threats, we constructed a test environment applying the \textit{HSS\textsubscript{A}} connected to an EPC controlled by the user organization, \textit{a VPN gateway}, and \textit{an external app server} as the baseline security elements, as shown in Figure~\ref{fig:AnaysisScope}. These are selected to make critical data controllable by the organization and protect the processes and data flow originating from the organization in terms of threat modeling \cite{DFD2007}. 

The HSS stores the cryptographic keys matched to the UEs. These keys are essential to protect the UEs and the network because they are used for mutual authentication between the UEs and the LTE network. The user organization needs to be able to control and protect such data even from the operating authority of the shared infrastructure. For this objective, the \textit{HSS\textsubscript{A}} independent from the HSS belonging to the shared infrastructure is added to the entities of the user organization. Because the authentication vector, which is generated from the keys and sent to the MME from the HSS for mutual authentication, does not contain the keys \cite{TS33.401, Cichonski2018NISTSecurity}, the objective can be achieved through this baseline security. A subscriber location function (SLF) is required to determine which HSS will authenticate when there are two or more HSSs \cite{TS23.228}. 

The LTE network provides the confidentiality and integrity protection mechanisms for control and user plane data. However, the application of the confidentiality protection of both planes remains an option of the network operator \cite{TS33.401, Cichonski2018NISTSecurity}. To protect the data flow originating from the user organization's IP networks, \textit{the VPN gateway} is required to encrypt the user plane communication channel between the UEs and the IP networks regardless of the operating authority. We used SSL VPN, which works on the application layer.

The last baseline security element is \textit{the external app server}. The user organization's internal web server providing the PPDR services is one of the assets belonging to the organization's legacy network where critical data are saved and transmitted. Although the radio frequency band is physically separated from those of commercial networks, it is risky to allow the UEs to connect to the web server directly. The app server is located between the web server and the UEs and operates as a proxy in the demilitarized zone (DMZ), which transmits traffic from the UEs to the web server in the proper format. We also implemented the service policy function in the web server, controlling the services to be provided to the app server. Through this mechanism, UEs allow only permitted services based on policy among the services running on the web server.

\subsection{Security Objectives}
In Table~\ref{tab:SecObjective}, we list the general security objectives for the data and the assets of the PPDR service environment linked to PS-LTE.

\begin{table}[H]
\caption{Security objective of the PPDR service system environment linked to PS-LTE}
\label{tab:SecObjective}
\centering
\begin{tabular}{cp{11.5cm}}
\toprule
\textbf{No.}	& \textbf{Security objectives}\\
\midrule
1 & Confidentiality, availability, and integrity of data between the user organization and its UEs should be protected. \\ 
2 & Unauthorized access to the assets of the user organization should be denied.\\
\bottomrule
\end{tabular}
\end{table}

Based on the first assumption described in section~\ref{sec:Assumptions}, we only set the confidentiality, availability, and integrity protection of the user plane data between the organization and the owned UEs as the first security objective. The second objective defines the denials of access that are unnecessary to provide and maintain LTE and PPDR services.

\subsection{Security Threat Analysis}
\label{sec:ThreatAnalysis}

For each security threat category listed in Table~\ref{tab:AnaysisScope}, we have drawn several potential threats. The number of threats and the threat examples are provided in Table~\ref{tab:ThreatCounts} \footnote{We leave out to list all threats due to the page limit and the security policy.}. Compared to \cite{DHS2017} and \cite{NISTIR8144}, which document security threats applicable to a general LTE and mobile environment, our analysis mostly targets the specific operational environment described in section~\ref{sec:EnvAnalysis}. 

\begin{table}[t]
\caption[Caption for LOF]{Numbers and examples of security threats in each category\footnotemark}
\label{tab:ThreatCounts}
\centering
\begin{tabular}{cccp{8cm}}
\toprule
\textbf{Category}	& \textbf{Label}	& \textbf{Count} & \textbf{Threat example}\\
\midrule
\multirow{3}{*}{UE}                                                   & 1.1   & 35     & device level information disclosure of type 1 UE; external data network connection of type 1 UE and the following threats.        \\ 
                                                                      & 1.2   & 7     & sending malicious attachments from type 2 to type 1 UE; other device connections to linked IP network through type 2 UE. \\
                                                                      & 1.3   & 6     & network level information disclosure of type 1 UE and the following spoofing threats.  \\ \midrule
eNodeB                                                                & 2.1   & 11     & external network connection of type 1 UE and the following information disclosure, denial of service, and elevation of privilege threats. \\ \midrule
EPC                                                                   & 3.1   & 15     & accessing the HSS\textsubscript{A} and the IP network by protocol not specified in standards; information disclosure of user plane data. \\ \midrule
\multirow{2}{*}{\begin{tabular}[c]{@{}c@{}}IP\\ network\end{tabular}} & 4.1   & 7     & VPN client identification spoofing; encrypted user plane data replay.\\
                                                                      & 4.2   & 9     & denial of service and elevation of privilege of the services running on web server. \\ \midrule
\multicolumn{2}{c}{\textbf{Total number}}                                    & \multicolumn{2}{c}{\textbf{90}} \\
\bottomrule
\end{tabular}
\end{table}

\footnotetext{All categories and labels are matched those in section~\ref{sec:AnalScope}.}

Table~\ref{tab:ThreatSeveriry} shows the statistics of the impact for each threat category. We assumed that the priority of system security is in order of integrity, confidentiality, and availability. In terms of data confidentiality, the leakage of the plain text user data and the system configuration data are more fatal than the ciphered data leakage. Therefore, we defined the categories of the threats impact \textit{crucial}, \textit{high}, \textit{medium}, and \textit{low}, reflecting the characteristics, and the definitions of each are described in Table~\ref{tab:ThreatSeveriry}.

It is worth noting that over 90\% of the threats belonging to the EPC category show crucial or high impact. Since all user and system data between the UE and the IP network, including the cryptographic setting information, are transferred through the S-GW and the P-GW, the threats in this category cause the most significant impacts.

\begin{table}[htb!]
\caption{Impact categories, the definitions, the count and the percentage by each threat categories}
\label{tab:ThreatSeveriry}
\begin{tabular}{ccccccc}
\toprule
\textbf{\begin{tabular}[c]{@{}c@{}}Impact \\category\end{tabular}}	& \textbf{Impact category definition} & \textbf{UE} & \textbf{eNodeB} & \textbf{EPC} & \textbf{IP Network} \\
\midrule
\textbf{Crucial} & the system integrity damage &  \begin{tabular}[c]{@{}c@{}}15 \\(31.2\%)\end{tabular}& \begin{tabular}[c]{@{}c@{}}3 \\(27.3\%)\end{tabular} &\begin{tabular}[c]{@{}c@{}}2 \\(26.7\%)\end{tabular} &
\begin{tabular}[c]{@{}c@{}}3 \\(18.8\%)\end{tabular}\\ \midrule
\textbf{High} & \begin{tabular}[c]{@{}c@{}}the confidentiality damage\\(deciphered data leakage)\end{tabular}& \begin{tabular}[c]{@{}c@{}}6 \\(12.5\%)\end{tabular} & \begin{tabular}[c]{@{}c@{}}2 \\(18.2\%)\end{tabular} & \begin{tabular}[c]{@{}c@{}}10 \\(66.7\%)\end{tabular} & \begin{tabular}[c]{@{}c@{}}5 \\(31.2\%)\end{tabular} \\\midrule
\textbf{Medium} & \begin{tabular}[c]{@{}c@{}}the confidentiality damage\\(system configuration data leakage)\end{tabular} &\begin{tabular}[c]{@{}c@{}}19 \\(39.6\%)\end{tabular}  & - & \begin{tabular}[c]{@{}c@{}}1 \\(6.6\%)\end{tabular} &\begin{tabular}[c]{@{}c@{}}6 \\(37.5\%)\end{tabular} \\ \midrule
\textbf{Low} & \begin{tabular}[c]{@{}c@{}}the confidentiality damage\\(ciphered data leakage) \\ the availability damage\end{tabular} & \begin{tabular}[c]{@{}c@{}}8 \\(16.7\%)\end{tabular} & \begin{tabular}[c]{@{}c@{}}6\\(54.5\%)\end{tabular}& \begin{tabular}[c]{@{}c@{}}2 \\(20\%)\end{tabular}&
\begin{tabular}[c]{@{}c@{}}2 \\(12.5\%)\end{tabular}\\
\bottomrule
\end{tabular}
\centering
\end{table}

\subsection{System Architecture and Security Requirements}
\label{sec:SysArchitecture}

\begin{figure}[hbt!]
    \centering
    \includegraphics[width=0.9\textwidth]{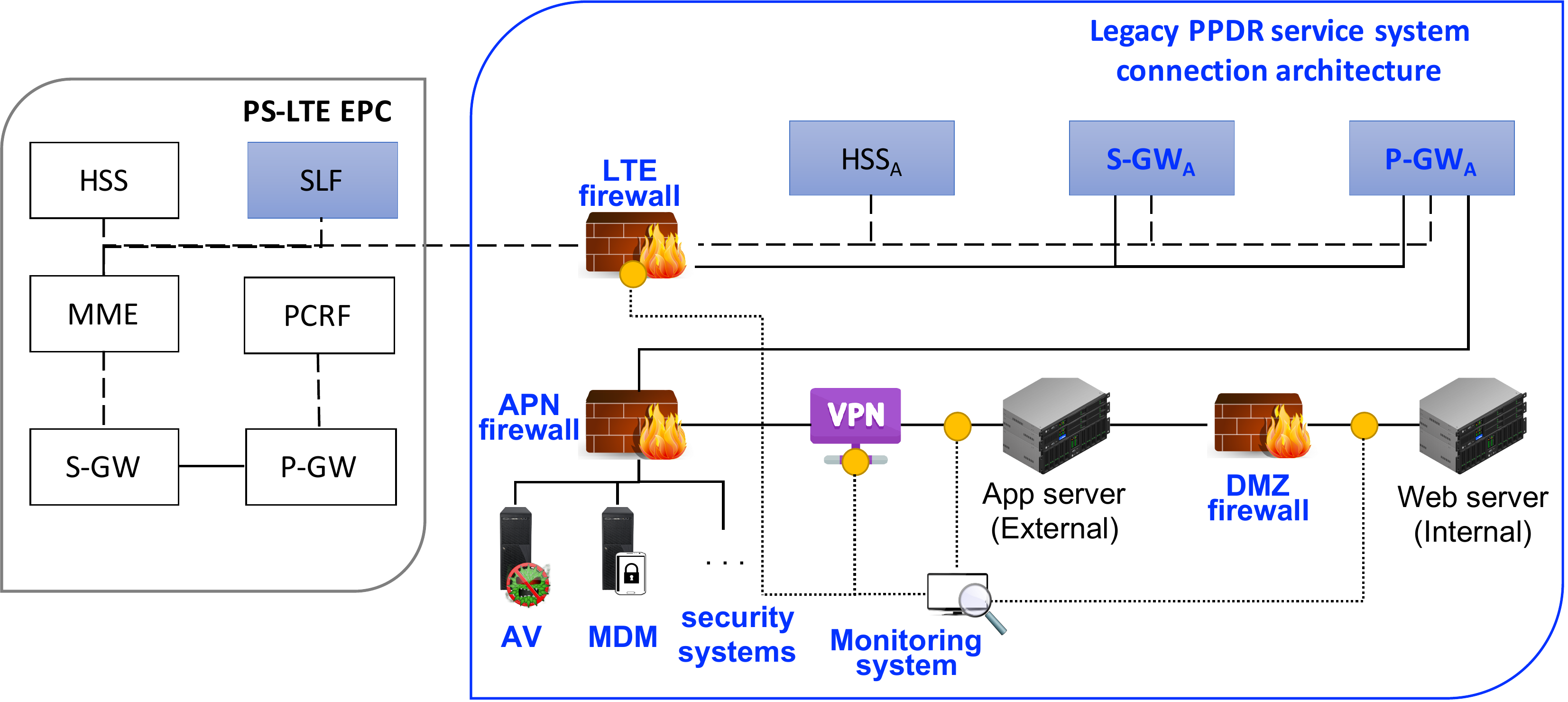}
    \caption{System architecture preventing security threats in the operation environment}
    \label{fig:SystemArchitecture}
\end{figure}

To prevent the analyzed security threats at the architecture level, we designed the system architecture (Figure~\ref{fig:SystemArchitecture}) by enhancing the security features of our test-bed structure (Figure~\ref{fig:AnaysisScope}).

The necessities of each entity except for those described in section~\ref{sec:testbedStructure} are as follows.

\begin{itemize}[label=$\bullet$]
    \item S/P-GW\textsubscript{A}: allocates IP to UEs within distinguishable range and protects from a sniffing of the user data 
    \item LTE firewall: enforces LTE signaling data transferred only between components through protocols defined in the specifications 
    \item APN firewall: protects IP network from unallowed external access and LTE components from unallowed internal access 
    \item DMZ firewall: allows network access only to legitimate app-web server pairs and certain services/protocols
    \item Security systems: enforce security policies to UEs
    \item Monitoring system: watches for prohibited or abnormal network access
\end{itemize}

The specifications of the security requirements used to protect from the threats, categorized as UEs, the infrastructure, and PPDR service system, are listed in Table~\ref{tab:SystemRequirements}. The UE category includes the requirements directly implemented in and applied to the UEs through security systems. The PS-LTE infrastructure category covers EPC, S/P-GW\textsubscript{A}, LTE, APN firewalls, and a transfer cipher function (VPN). The PPDR service system category corresponds to apps and web servers, as well as a DMZ firewall.

\section{Conclusion and Future Work}
\label{sec:Conclusion}
To design a security model in a PPDR operation environment under the PS-LTE infrastructure, we first introduced a framework for designing the security model for the environment under which the technologies are adopted for a specific purpose. In addition, we demonstrated the application results in the target environment. As a result, the system architecture and the security requirements for the system are designed as the security model. 

The main observation in the framework application example in section ~\ref{sec:Application} is that even the cryptographic setting information can be sniffed from the S-GW and the P-GW. For a reason, we proposed constructing the S-GW and the P-GW that are owned by the user organization. However, the security objective can also be achieved by using IPSec VPN, which works on the network layer. Since IPSec VPN supports the symmetric cipher, the cryptographic key exchange is not required.

During the proposed framework application under the PPDR operational environment, the PS-LTE technologies defined in standards such as GCSE, ProSe, and IOPS are excluded from the analysis based on certain assumptions. However, these technologies may cause security threats, which have yet to be researched. We would like to extend this analysis's scope to increase the security of the PPDR operational environment.

\section{Acknowledgement}
The study was funded by Institute for Information and communications Technology Promotion (Grant No. 2020-0-00374, Development of Security Primitives for Unmanned Vehicles).

\begin{table}[t]
\caption{Security requirements}
\label{tab:SystemRequirements}
\centering
\begin{tabular}{p{2cm}p{10cm}}
\toprule
\textbf{Category}            & \textbf{Security requirements} \\\midrule
\multirow{15}{*}{UE} & \hangindent=1.5mm - restrict to only access the LTE network (PLMN\footnotemark) that the shared infrastructure (PS-LTE) operates \\
                    & \hangindent=1.5mm - prohibit from accessing other IP networks except those operated by the owner organization \\
                    & \hangindent=1.5mm - restrict not to provide network functionalities that enable other devices to access the IP networks, e.g., hotspot and tethering \\
                    & \hangindent=1.5mm - protect external storages to be read \\
                    & \hangindent=1.5mm - apply data leakage protection, e.g., use data store ciphering or build a cloud system to prohibit UE from storing data  \\
                    & \hangindent=1.5mm - enforce memory protection and apply PIN to USIM \\
                    & \hangindent=1.5mm - enforce user to UE authentication \\
                & \hangindent=1.5mm - allow to transceive only the PS-LTE radio frequency bands (enforced requirement) \\
                &  \hangindent=1.5mm - enable minimum functionality of UE when network is disconnected or the mobile device management (security) policies are not applicable \\
                & \hangindent=1.5mm - allow only the mobile service application to be installed in the white list \\
                & \hangindent=1.5mm - enforce encryption/decryption of all data tranceived from/to UE\\
                &  \hangindent=1.5mm - enforce the security policies to be applied after the factory initialization of UE\\
                &  \hangindent=1.5mm - protect the mobile applications to enforce security policies to be terminated and removed\\
                 & \hangindent=1.5mm - keep the versions of OS and the mobile applications installed in UE up to date and confirm the integrity of the update files\\
                 & \hangindent=1.5mm - prohibit to execute all functionalities of rooted UEs\\
\hline
\multirow{8}{*}{\begin{tabular}[c]{@{}c@{}}PS-LTE   \\infrastructure\end{tabular}}      & \hangindent=1.5mm - enforce multi-factor authentication for user to UE, user to infrastructure (network), and user to services authentications\\
                    & \hangindent=1.5mm - check validity of IMSI and IMEI pair and user and UE pair during network connection\\
                    & \hangindent=1.5mm - allow connections between LTE components only specified in standards and restrict the connections to service/protocol level  \\
                    & \hangindent=1.5mm - allow UE connection to IP network only to those allocated IP within distinguishable range \\
                    & \hangindent=1.5mm - continuously change ciphering keys for transferring data even within the same session\\
                    & \hangindent=1.5mm - enforce the traffic tranceived between type 1 UE and IP network of an organization to pass P-GW\textsubscript{A} and S-GW\textsubscript{A}, not P-GW nor S-GW \\
                    & \hangindent=1.5mm - allocate IP address to type 1 UEs distinct to other UEs \\
                    &\hangindent=1.5mm - use security certified devices consisting the security systems\\
\hline
\multirow{7}{*}{\begin{tabular}[c]{@{}c@{}}PPDR service  \\system\end{tabular}}      &\hangindent=1.5mm - provide API to call functions in web server and defines/sets authorization levels considering user types \\
                    &\hangindent=1.5mm - prohibit from executing the functions for which the API is not defined \\
                    &\hangindent=1.5mm -provide services run in web server to UEs only through app server \\
                    &\hangindent=1.5mm - do not store any generated or passing data to app server during service; app server behaves like a proxy \\
                    &\hangindent=1.5mm - develop the mobile service applications as in-app fashion and check if requests sent from UE generated by the applications; apps must not rely on browser\\    
                    &\hangindent=1.5mm - develop mobile service applications applying obfuscation technologies \\
                    &\hangindent=1.5mm - develop the app/web server programs and the mobile service applications following the secure coding norms\\
                    &\hangindent=1.5mm - use security certified devices consisting the security systems\\
\bottomrule
\end{tabular}
\end{table}

\footnotetext{Public Land Mobile Network}

\bibliographystyle{splncs04}
\bibliography{Mendeley,Reference}

\end{document}